\documentclass[12pt,aps,prl,amsmath,amssymb,sort&compress,comma,super,nofootinbib,floatfix]{revtex4-1}
\usepackage{graphicx}
\usepackage{epstopdf}
\usepackage{subfigure}  
\begin{document}

\newcommand{\etal}{{\it et al.}\/}
\newcommand{\gtwid}{\mathrel{\raise.3ex\hbox{$>$\kern-.75em\lower1ex\hbox{$\sim$}}}}
\newcommand{\ltwid}{\mathrel{\raise.3ex\hbox{$<$\kern-.75em\lower1ex\hbox{$\sim$}}}}

\title{Dependence of $T_c$ on the $q-\omega$ structure of the spin-fluctuation spectrum}

\author{Thomas Dahm}
\affiliation{Universit\"at Bielefeld, Fakult\"at f\"ur Physik, Postfach 100131, D-33501 Bielefeld, Germany}

\author{D.J.~Scalapino}
\affiliation{Department of Physics, University of California, Santa Barbara, CA 93106-9530, USA}


\begin{abstract}
A phenomenological spin-fluctuation analysis \cite{ref:1}, based upon inelastic
neutron scattering (INS) and angular resolved photoemission spectroscopy (ARPES) data for
${\rm YBCO}_{6.6}(T_c=61K)$, is used to calculate the functional derivative of
the d-wave eigenvalue $\lambda_d$ of the linearized gap equation with respect
to the imaginary part of the spin susceptibility
$\chi''(q,\omega)$ at 70K. For temperatures near $T_c$, the variation of $T_c$
with respect to $\chi''(q,\omega)$ is proportional to this functional derivative.
Based on this, we discuss how different parts of the $q$ and $\omega$ dependent
spin-fluctuation spectrum of YBCO$_{6.6}$ contribute to $T_c$.
\end{abstract}


\maketitle


For the traditional electron-phonon driven superconductors, the Eliashberg theory for
the transition temperature $T_c$ depends upon the spectral function of the
effective interaction $\alpha^2F(\omega)$ due to the exchange of phonons and the
Coulomb pseudo potential $\mu^*$ \cite{ref:2,ref:3}. Electron tunneling
measurements provided experimental results for these quantities which were used
to calculate $T_c$ \cite{ref:4}. Questions then arose as to how the different
frequency regions contributed to $T_c$. In order to understand this, Bergmann
and Rainer \cite{ref:5} used electron tunneling data 
and calculated the functional derivative of $T_c$ with respect to $\alpha^2F(\omega)$.
They found that while all parts of the phonon spectrum contributed to $T_c$,
$\delta T_c/\delta\alpha^2F(\omega)$ peaked for $\omega\sim7 T_c$, falling off
at higher frequencies. In contrast to the electron-phonon case, the pairing interaction
in the cuprates has an important momentum dependence so that one would like
to understand how both different q and $\omega$ regions  contribute to $T_c$.
While one lacks an equivalent Eliashberg theory,
  fluctuation exchange
(FLEX) calculations of the variation of $T_c$ with changes in the $q-\omega$ spin-fluctuation
spectral weight for the 2D Hubbard model have been reported \cite{ref:6}. 
Here we take a more
phenomenological approach and explore how inelastic neutron scattering (INS)
data \cite{ref:7,ref:8,ref:9,ref:10} for the dynamic spin susceptibility $\chi''(q,\omega)$ along with angular
resolved photoemission (ARPES) results \cite{ref:11,ref:12} for ${\rm YBCO}_{6.6}$
can provide insight into how different parts of the $q$ and $\omega$ dependent
spin-fluctuation spectrum contribute to the pairing. We find for
${\rm YBCO}_{6.6}$ that there is pair breaking for $\omega\ltwid $ 25 meV and that the
dominant pairing strength comes from the upper branch of the ${\rm YBCO}_{6.6}$
spin-fluctuation spectrum.

\begin{figure}[htbp]
  \includegraphics[width=7.5cm]{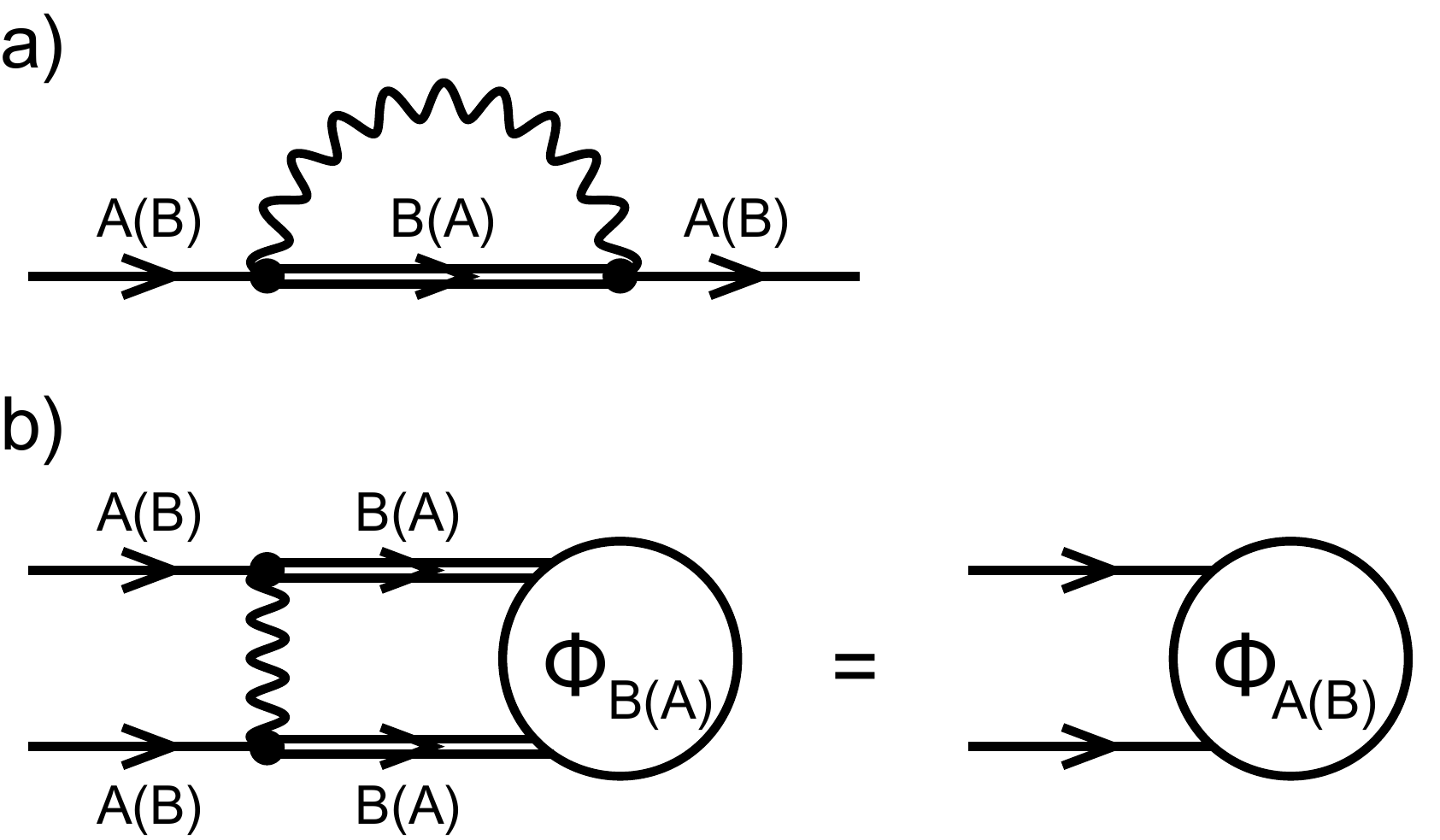}
\caption{a) One loop self-energy diagram. Here the wiggly line represents the
interaction, Eq.~(\protect\ref{eq:1}), and the solid line represents the dressed
single particle Green's function.; b) The Bethe-Salpeter linearized gap diagram.\label{fig:1}}
\end{figure}

Within the spin-fluctuation framework, the diagrams for the one-electron self-energy
and the linearized gap equation are shown in Figure~\ref{fig:1}.  Here the wiggly
line represents the effective interaction
\begin{equation}
  V_{\rm eff}(q,\omega)=\frac{3}{2}\bar U^2\chi(q,\omega)
\label{eq:1}
\end{equation}
with $\chi(q,\omega)$ the $q$ and $\omega$ dependent spin susceptibility
measured by inelastic neutron scattering \cite{ref:7,ref:8,ref:9,ref:10}. Here, 
as in Ref. [1], the parameterized form for $\chi(q,\omega)$ that we will use
describes the odd symmetry channel with respect to the interchange
of adjacent $ CuO_2$ bilayers. This is the channel that contains the spin resonance 
and the one whose q and $\omega$ contributions to the pairing  we
will examine.
The
solid lines represent the one-electron Green's function $G(k,\omega)$ with the
 one-loop self-consistent
self-energy illustrated in Fig.~\ref{fig:1}a. In our calculations, the imaginary
parts of the one-loop self-energies for the antibonding (A) and bonding (B)
two-layer bands are given by
\begin{equation}
  {\rm Im}\Sigma_{\rm A,B}(k,\omega)=\frac{1}{N}\sum_Q\int^\infty_{-\infty}
	\frac{d\Omega}{\pi}[n(\Omega)+f(\Omega-\omega)]{\rm Im} V_{\rm eff}(Q,\Omega)
	{\rm Im}{\rm G}_{\rm B,A}(k-Q,\omega-\Omega)
\label{eq:2}
\end{equation}
Here $n$ and $f$ are the Bose and Fermi functions, respectively, and $Q$ is
summed over the 2D Brillouin zone. The odd symmetry of $\chi(q,\omega)$ channel couples  $\Sigma_{\rm A}$
to ${\rm G}_{\rm B}$ and $\Sigma_{\rm B}$ to ${\rm G}_{\rm A}$. 
 The real parts of $\Sigma_{\rm A,B}$ are evaluated by a Kramers-Kronig
transformation.

The dispersion relations of the unrenormalized electron bands $\varepsilon^{\rm A,B}(k)$
of the two-layer YBa$_2$Cu$_3$O$_{6.6}$ system are modeled by tight binding
parameters and a chemical potential. In the iterative calculation of the self-energy,
these parameters are adjusted to preserve the observed ARPES bonding and
antibonding Fermi surfaces of the dressed electrons. The coupling $\bar U$ was chosen to fit
the observed  nodal Fermi velocity \cite{ref:1}, but its precise magnitude plays a negligible role
in the  q and $\omega$ dependence of the functional derivative.

The imaginary part of the linearized gap equation, Fig.~\ref{fig:1}b, is given by
\begin{eqnarray}
  \frac{1}{\pi N}\sum_{k'}\int^\infty_{-\infty} \nonumber
	&&d\omega'[n(\omega-\omega')+f(-\omega')]\ {\rm Im}V_{\rm eff}(k-k',\omega-\omega')\\ \nonumber
	&&{\rm Im}\left(
  	\frac{\phi_{\rm B,A}(k',\omega')}
	    {(\omega'Z_{\rm BA})^2-(\tilde\varepsilon^{\rm B,A}(k))^2}
		\right)=\lambda_d(T){\rm Im}\phi_{\rm A,B}(k,\omega)\\
\label{eq:3}
\end{eqnarray}
with
\begin{equation}
  \omega Z_{\rm B,A}(k,\omega)=
	\omega-\frac{1}{2}\left(\Sigma_{\rm B,A}(k,\omega)-\Sigma^*_{\rm B,A}(k,-\omega)\right)
\label{eq:4}
\end{equation}
and
\begin{equation}
  \tilde\varepsilon^{\rm B,A}(k)=\varepsilon^{\rm B,A}(k)+\frac{1}{2}
	\left(\Sigma_{\rm B,A}(k,\omega)+\Sigma^*_{\rm B,A}(k,-\omega)\right)
\label{eq:5}
\end{equation}
The eigenfunction of Eq.~(\ref{eq:3}) with the largest low temperature eigenvalue
has $d$-wave symmetry and its eigenvalue $\lambda_d(T)$ approaches 1 as $T$ goes
to $T_c$. In the following we will calculate the functional derivative of
$\lambda_d(T)$ with respect to ${\rm Im}\chi(q,\omega)$ for momentum $q$ along the
diagonal of the Brillouin zone, using INS results for YBCO$_{6.6}$
measured at $T=70$K. The $T_c$ of YBa$_2$Cu$_3$O$_{6.6}$ is 61K and for $T$ near
$T_c$, the variation of $T_c$ with respect to ${\rm Im}\chi(q,\omega)$ is
proportional to $\delta\lambda_d/\delta{\rm Im}\chi(q,\omega)$. To calculate
$\delta\lambda_d/\delta{\rm Im}\chi(q,\omega)$ at $\omega_0$ and $q_0$, we
set ${\rm Im}\tilde\chi(q,\omega)={\rm Im}\chi(q,\omega)+a\delta(\omega-\omega_0)
\delta(q-q_0)$ and numerically evaluated $(\tilde\lambda_d-\lambda_d)/a$. Here
$a=0.1$ is small compared to the integrated spectral weight over a phase space
region $\Delta q\Delta\omega$ with $\frac{\Delta q}{q}=\frac{\Delta\omega}{\omega}=0.01$.

\begin{figure}[thbp]
\begin{center}
\subfigure{
\includegraphics[width=7.5cm,angle=270]{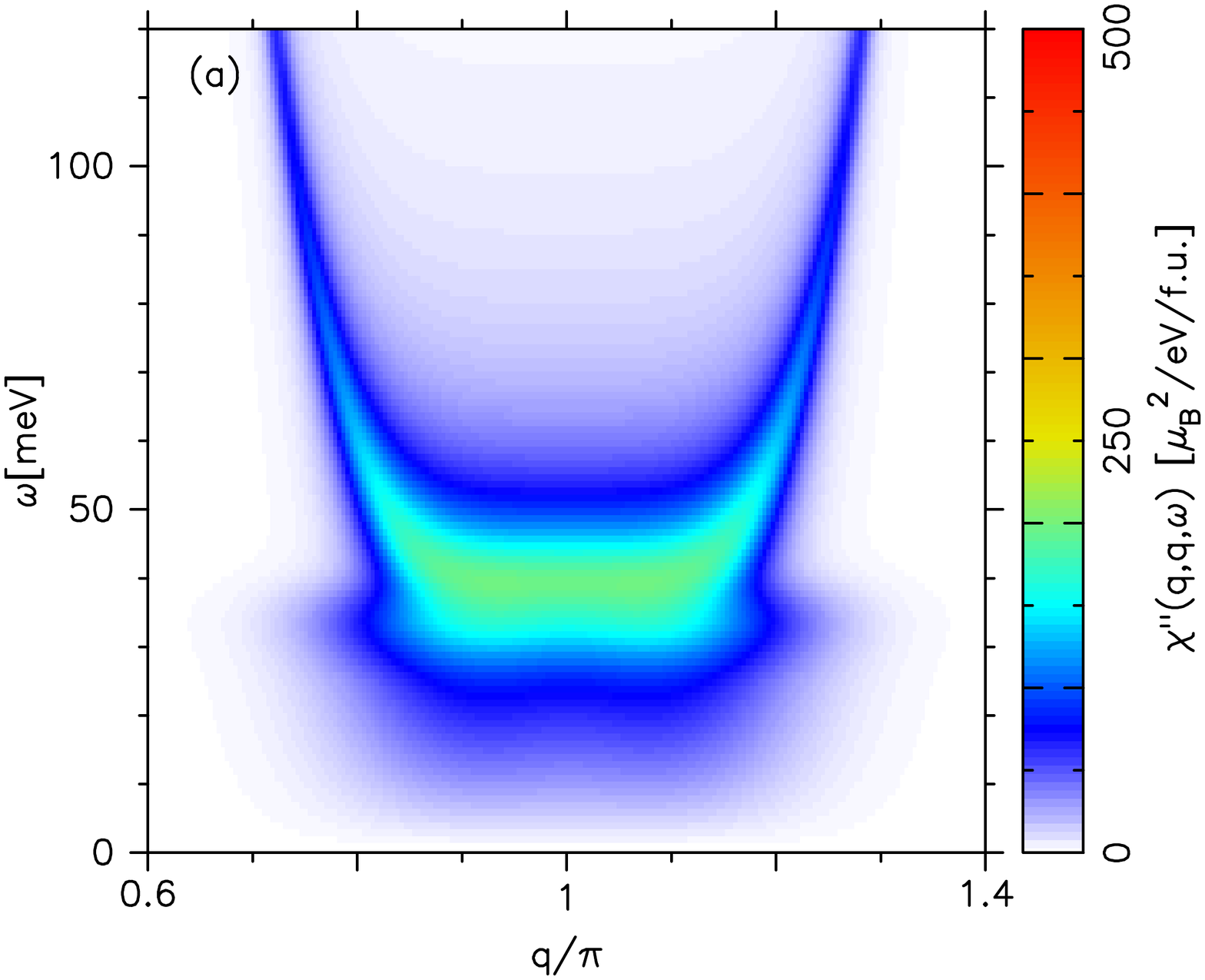}}
\subfigure{
\includegraphics[width=7.5cm,angle=270]{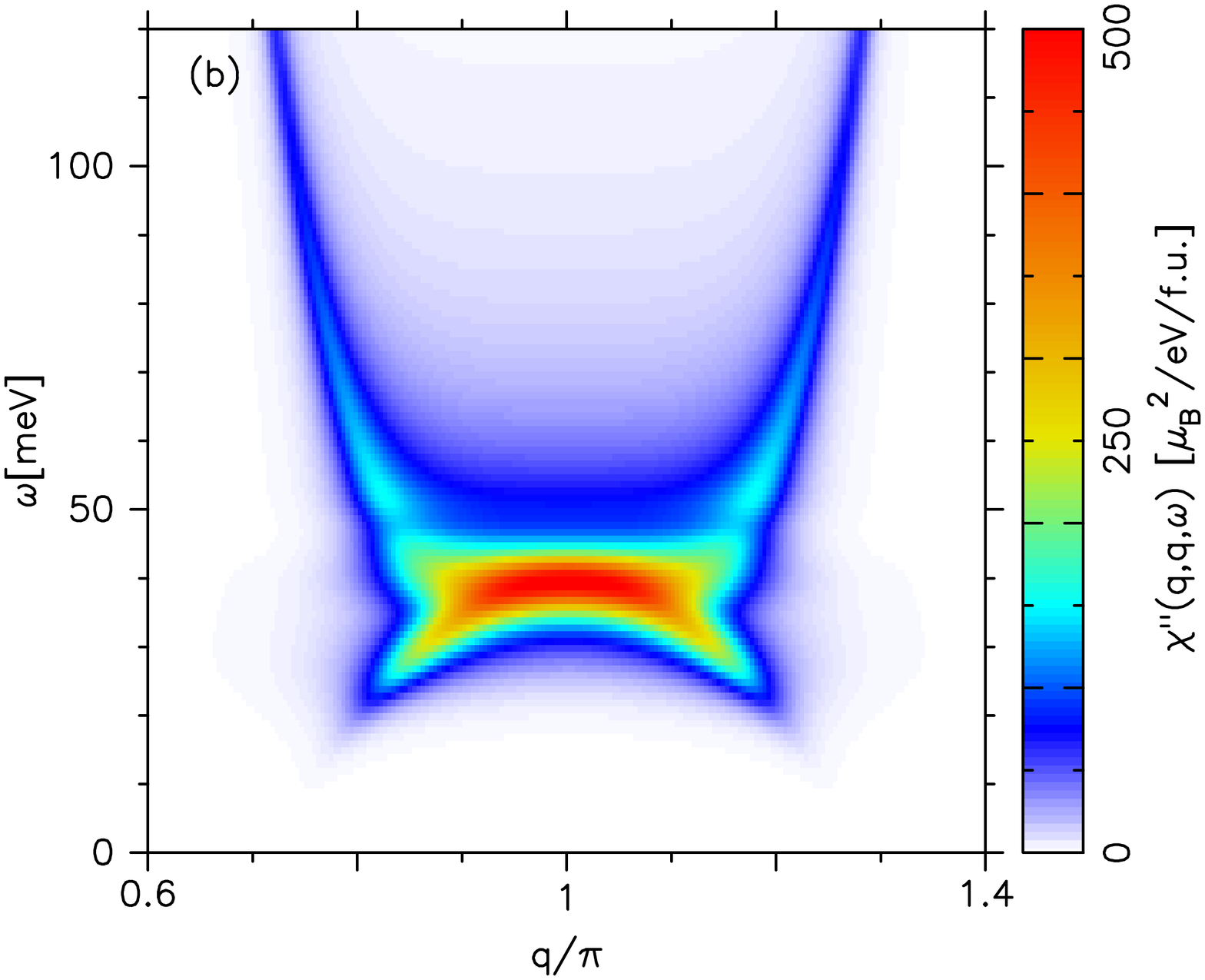}}
\caption{(a) Parameterization of $\chi''(q,\omega,70{\rm K})$ obtained from inelastic
neutron scattering on YBCO$_{6.6}$ at $T=70$K (see supplementary material of
Ref.~\protect\cite{ref:1}). Here $q$ runs along the diagonal direction of the
Brillouin zone, i.e. $q_x=q_y=q$. (b) A similar plot of $\chi''(q,\omega,5{\rm K})$ for $T=5$K.
\label{fig:2}}
\end{center}
\end{figure}

\begin{figure}[htbp]
\begin{center}
\subfigure{
\includegraphics[width=8.0cm,angle=270]{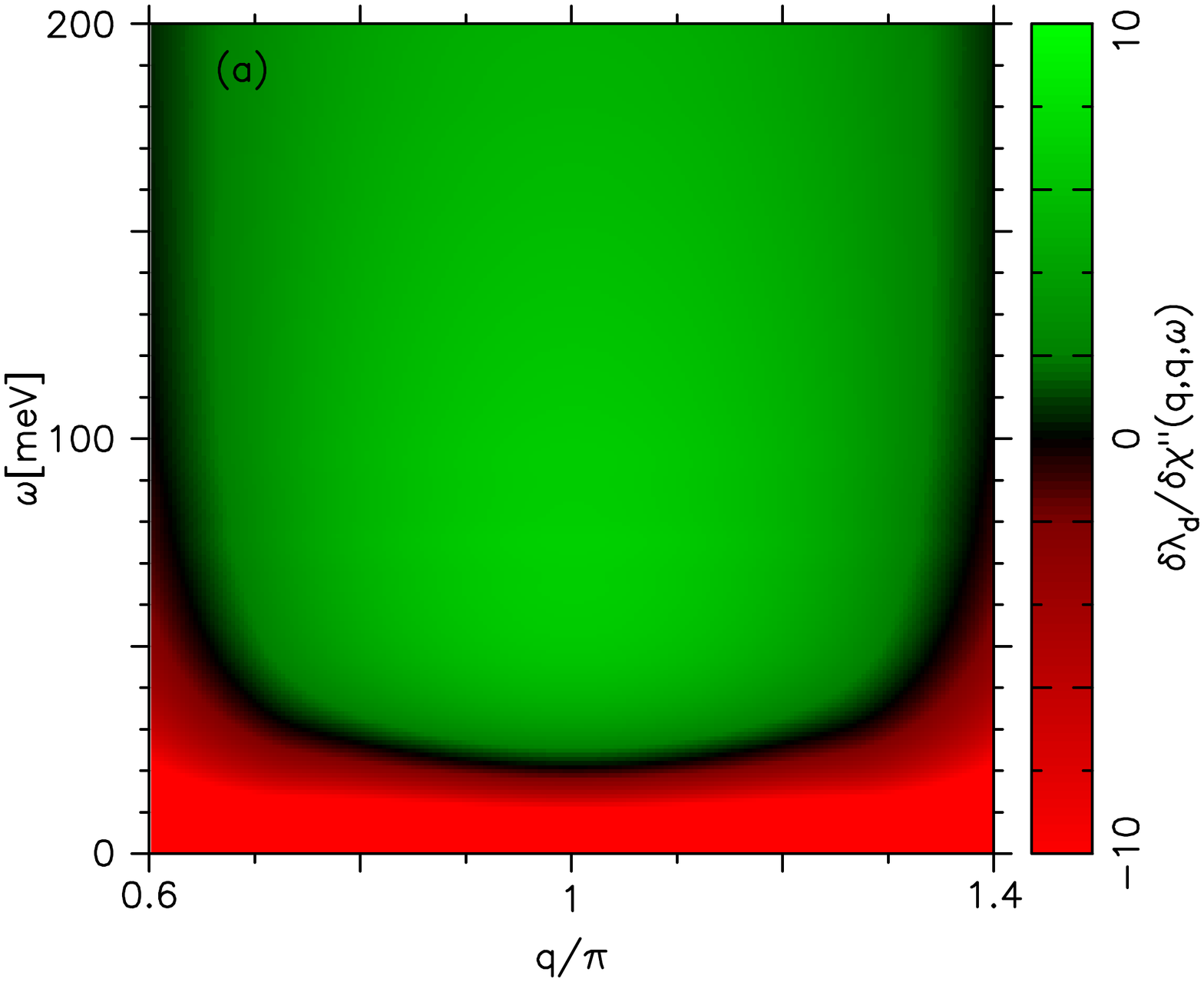}}
\subfigure{
\includegraphics[width=8.0cm,angle=270]{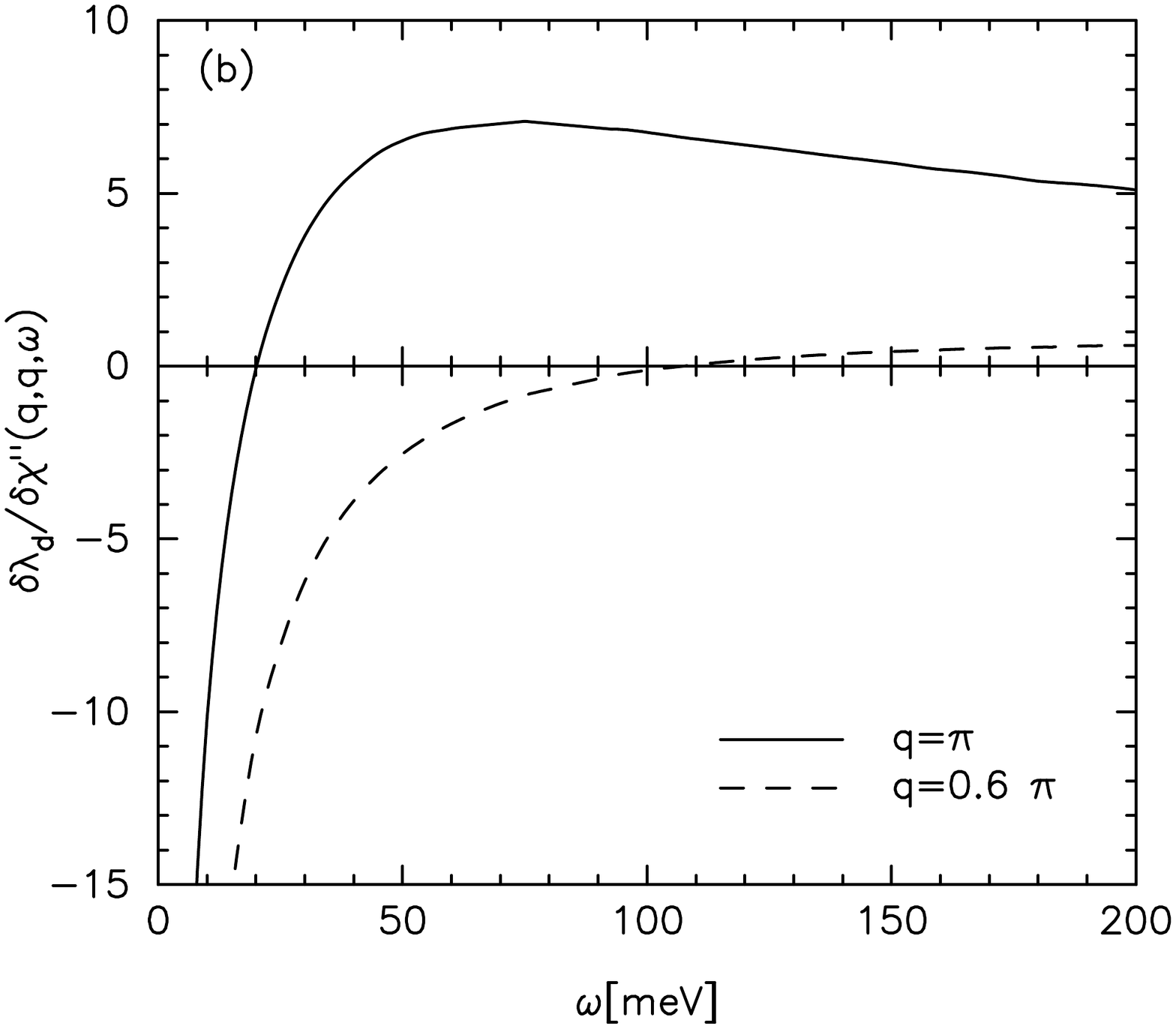}}
\caption{(a) The functional derivative of the d-wave eigenvalue $\lambda_d$ with
	respect to $\chi''(q,\omega,70{\rm K})$ versus $\omega$ and $(q_x=q,q_y=q)$.
	The normalization is such that
	$\chi''(q,\omega,70{\rm K})\frac{\delta\lambda_d}{\delta\chi''(q,\omega,70{\rm K})}$
	averaged over the $q-\omega$ phase space shown in the figure is 1. (b) The $\omega$ dependence of $\frac{\delta\lambda_d}{\delta\chi''(q,\omega,70{\rm K})}$
	for $q=\pi$ (solid) and $q=0.6\pi$ (dashed).
\label{fig:3}}
\end{center}
\end{figure}

A plot of a parameterized fit \cite{ref:1} of the INS data showing $\chi''(q,\omega,70{\rm K})$
for YBCO$_{6.6}$ at $T=70$K is shown in Fig.~\ref{fig:2}a.
For this underdoped cuprate there is a clear pseudogap and the spin-fluctuation
spectrum can be considered as having upper and lower branches. In Fig.~\ref{fig:2}b
a similar  plot of $\chi''(q,\omega,5{\rm K})$ for $T=5$K shows the development of
the hourglass dispersion and spin resonances in the superconducting state. The
functional derivative $\delta\lambda_d/\delta({\rm Im}\chi(q,\omega,70K))$
plotted in Fig.~\ref{fig:3} provides a map showing how different regions of the
$q-\omega$ phase space contribute to the pairing. The strongest pairing occurs 
for large momentum transfers with frequencies extending from $\sim$40meV to several hundred meV.  At high frequencies for q along the diagonal, the functional derivative varies as $-\cos(q) / \omega$.
At frequencies lower than $\sim$25meV, adding additional spin-fluctuation spectral weight reduces $\lambda_d $ corresponding to a suppression of $T_c$.
 This reflects the well known pair breaking effects of low frequency spin fluctuations \cite{ref:13}. As opposed to the phonon case where fluctuations at 
 all frequencies contribute to $T_c$, the low frequency spin fluctuations act
 as static magnetic impurities and suppress $T_c$. 
 At small momentum transfers, the spin-fluctuations predominantely scatter pairs between regions of the Fermi surface
  where the d-wave gap has the same sign and the functional derivative becomes negative.  As seen from the dashed curve of Fig3b for $q = (0.6\pi,0.6\pi)$, even at large 
  frequencies this effect of the d-wave form factor leads to negligible pairing. Overall it is clear that the dominant 
  contribution to the pairing is coming from the upper branch of the spin fluctuations.


\begin{figure}[htbp]
\begin{center}
\subfigure{
\includegraphics[width=7.5cm,angle=270]{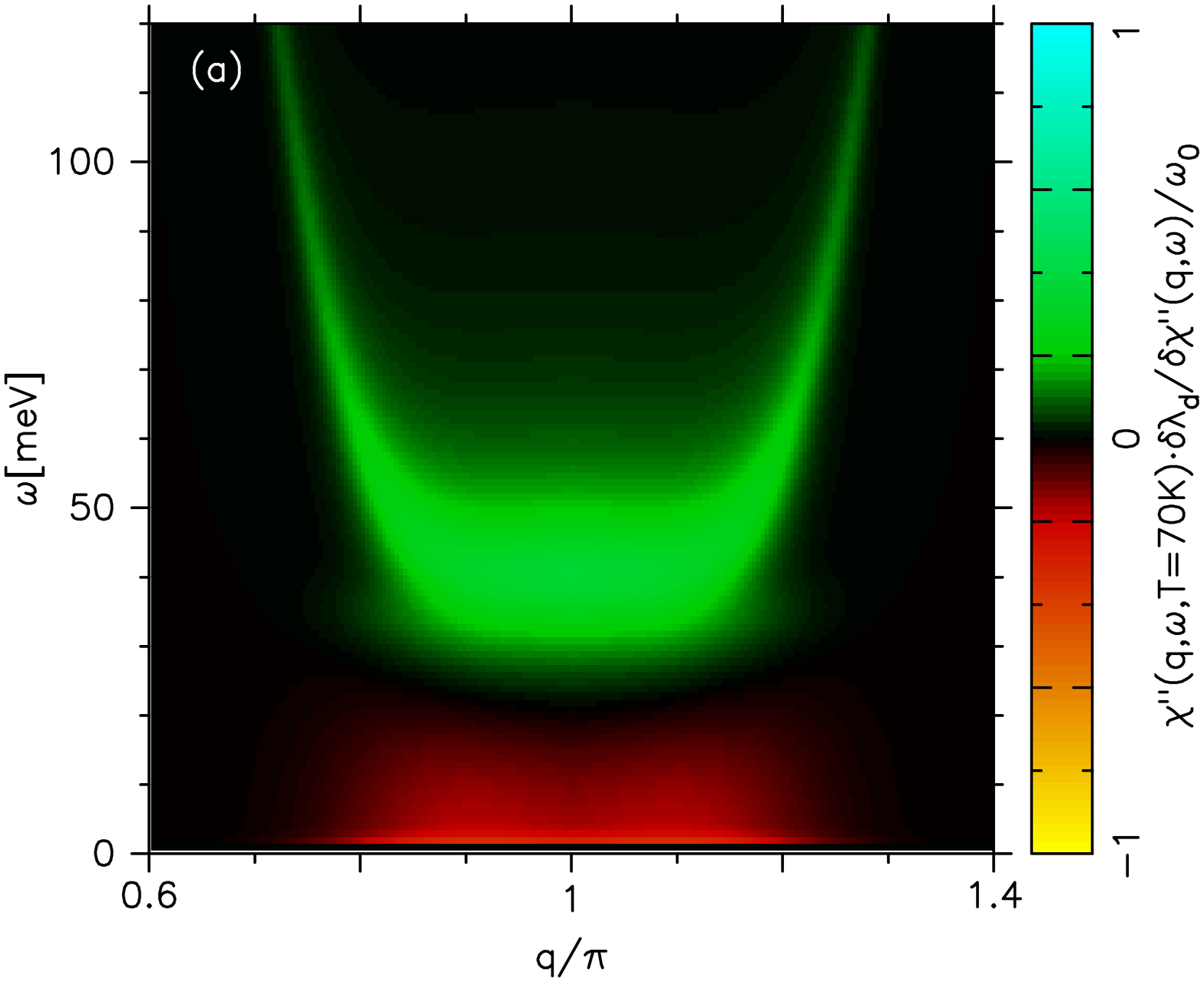}}
\subfigure{
\includegraphics[width=7.5cm,angle=270]{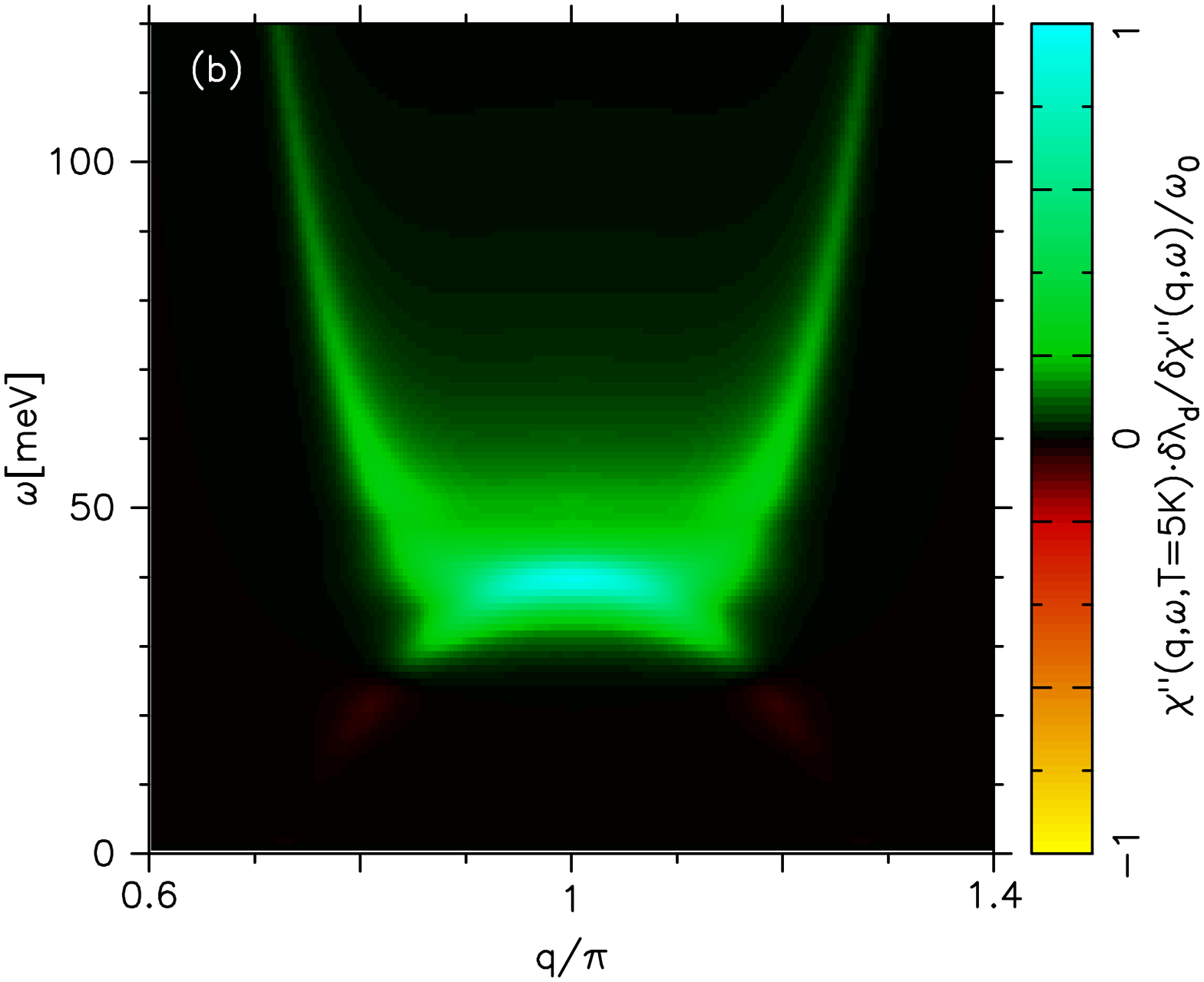}}
\caption{(a) Plot of $\chi''(q,\omega,70{\rm K})$ times the functional derivative
shown in Fig.~\protect{\ref{fig:3}} divided by $0.8\pi\omega_0$. This quantity
gives the contribution to the d-wave eigenvalue $\lambda_d$ from a $\Delta q\times\Delta\omega$
region of $(q,\omega)$ phase space for $T=70$K. (b) A similar plot using the
functional derivative computed at $T=70$K and $\chi''(q,\omega,T=5{\rm K})$. These
plots illustrate how different parts of the spin-fluctuation spectrum of YBCO$_{6.6}$
contribute to $\lambda_d$. There is an increase of the pairing strength that
occurs as $T$ drops below $T_c$ and weight in the spin-fluctuation spectrum shifts
to higher frequencies.\label{fig:4}}
\end{center}
\end{figure}

The product of $\chi''(q,\omega,70{\rm K})$ times the functional derivative
$\delta\lambda_d/\delta\chi''(q,\omega)$ is shown in Fig.~\ref{fig:4}a.
This quantity illustrates how different parts of the spectrum contribute to the
pairing. The red pairbreaking region below $\sim25$~meV in the functional derivative shown in
Fig.~\ref{fig:3} is not so destructive for superconductivity in YBCO$_{6.6}$
because there is not so much weight in $\chi''(q,\omega,70{\rm K})$ in this frequency region.
In this sense the opening of the pseudogap in $\chi''(q,\omega)$ enhances the
pairing. The green region of the functional derivative in Fig.~\ref{fig:3} is
emphasized by the ``upper branch" of the spin-fluctuation spectrum. It is
interesting to examine a similar plot in which the functional derivative of
Fig.~\ref{fig:3} is multiplied by the $\chi''(q,\omega,T=5{\rm K})$. As seen in
Fig.~\ref{fig:2}b, when $T$ decreases from 70K to 5K, the low frequency part
of the spin-fluctuation spectral weight decreases and the intensity increases
in regions that contribute to the pairing. This behavior is clearly reflected in
Fig.~\ref{fig:4}b. As noted in Ref.~\cite{ref:1}, if Im$\chi(q,\omega,70K)$ is
replaced by the 5K INS data Im$\chi(q,\omega,5K)$ in the Bethe-Salpeter
equation, one finds an approximate 50\% increase in $\lambda_d$.

Central to the proposal that antiferromagnetic spin-fluctuations provide the
pairing interaction responsible for superconductivity in the cuprate
superconductors is the idea that there is a significant coupling between 
the spins  and the doped holes.
A consequence of this is that when the system
becomes superconducting and a d-wave gap opens in the  quasi-particle
spectrum,the anti-ferromagnetic spin-fluctuation spectrum will be altered.
In the superconducting state, spin-fluctuation spectral weight is shifted
up in frequency along with the formation of a spin resonance changing the strength
of the pairing interaction. The results shown in Fig.~\ref{fig:4}
provide evidence that the pairing
strength increases below $T_c$ as spin-fluctuation spectral weight is removed
from low frequencies and shifted to frequencies of order $2\Delta_{\rm max}$.
One consequence of this is that the magnitude of the d-wave gap will increase
more rapidly as $T$ drops below $T_c$ and 2$\Delta_{max}/kT_c$ will be larger
than found from predictions based upon a pairing interaction which remains the same
below $T_c$. 

\section*{Acknowledgments}

The authors would like to thank Steve Kivelson and John Tranquada for helpful discussions. DJS was
supported by the SciDAC program of the U.S. Department of Energy Division of
Materials Sciences and Engineering. 



\end{document}